\documentclass[12pt]{iopart}

\usepackage{graphicx}
\usepackage{dcolumn}
\usepackage{bm}
\usepackage{rotating}

\begin{document}

\title{Electronic properties of Francium diatomic compounds and prospects for cold molecule formation}

\author{M.~Aymar}
\address{Laboratoire~Aim\'{e}~Cotton, CNRS, B\^{a}timent~505, Universit\'{e}~d'Orsay, 91405~Orsay-Cedex, France}
\author{O.~Dulieu}
\ead{olivier.dulieu@lac.u-psud.fr}
\address{Laboratoire~Aim\'{e}~Cotton, CNRS, B\^{a}timent~505, Universit\'{e}~d'Orsay, 91405~Orsay-Cedex, France}
\author{F. Spiegelman}
\address{Laboratoire de Chimie et Physique Quantiques, UMR 5626, IRSAMC, Universit\'{e} Paul Sabatier, 118 route de Narbonne, F-31062 Toulouse, France }

\date{\today}

\begin{abstract}
In this work we investigate the possibility to create cold Fr$_2$, RbFr, and CsFr molecules
through photoassociation of cold atoms.
Potential curves, permanent and transition dipole moments for the Francium dimer and for the RbFr
and RbCs molecules are determined for the first time. The Francium atom is modelled as a one valence
electron moving in the field of the Fr$^+$ core, which is described by a new pseudopotential with
averaged relativistic effects, and including effective core polarization potential. The molecular 
calculations are performed for both the ionic species Fr$_2^+$, RbFr$^+$, CsFr$^+$ and the corresponding
neutral, through the CIPSI quantum chemistry package where we used new extended gaussian basis sets
for Rb, Cs, and Fr atoms. As no experimental data is available, we discuss our results
by comparison with the Rb$_2$, Cs$_2$, and RbCs systems. The dipole moment of CsFr reveals an electron transfer
yielding a Cs$^+$Fr$^-$ arrangement, while in all other mixed alkali pairs the electron is transferred
towards the lighter species. Finally the perturbative treatment of the spin-orbit coupling at large distances
predicts that in contrast with Rb$_2$ and Cs$_2$, no double-well excited potential should be present in Fr$_2$,
probably preventing an efficient formation of cold dimers via photoassociation of cold Francium atoms.
\end{abstract}


\maketitle

\section{Introduction}
\label{intro}

As discussed for instance by Sprouse {\it et al} \cite{sprouse2002} the recent development
of laser cooling and trapping of  radioactive atoms opens ways for new
investigations like the search for the electric dipole moment (EDM), $\beta$ decay, 
cold atom-atom collisions, Bose-Einstein condensation and more precise atomic clocks.
As the heaviest alkali-metal atom, Francium offers a sensitive test of the influence 
of relativistic and quantum electrodynamic effects (QED) on the atomic structure. It
is also an excellent system to study weak interactions in atoms such as atomic 
parity non-conservation (PNC) \cite{behr1993,bouchiat1997}.

However, none of the Francium istopes is stable. The longest-lived isotope is $^{223}$Fr
with a half-time of 21.8 min, while the shortest-lived isotope
$^{215}$Fr has a half-time of 0.12 $\mu$s. Several groups have been successful in trapping
long-lived isotopes in a magneto-optical trap (MOT): in 1995,
Simsarian {\it et al} captured $^{210}$Fr atoms (half-time 3.2 min) produced at
the Stony Brook LINAC \cite{simsarian1996a,sprouse1998}; a collaboration between the 
groups of C. Wieman from the University of Colorado and  
H. Gould from Lawrence Berkeley Laboratory lead to the trapping
 of $^{221}$Fr (half time 4.8 min) \cite{lu1997}; more recently, a $^{210}$Fr MOT has been  set up
in Legnaro (Italy) \cite{atutov2003},  but to our knoweledge its optimization is still under way.

Even if the spectroscopy of Francium atom is not yet fully investigated, a noticeable 
amount of experimental and theoretical data is now available for energy levels,
hyperfine splittings, and transition dipole moments. In contrast, 
experimental conditions allowing to form Francium dimers have not yet been achieved, 
and the availability of cold Francium atoms in MOT's may allow one day to perform the
photoassociation spectroscopy of such radioactive molecules. The creation of a
mixed species cold atom trap involving Francium and Rubidium or Cesium appears as a
closer possibility \cite{orozco2005}, which would allow to explore the RbFr or CsFr
heteronuclear molecules. 

On the theoretical side, no potential curves have been published
yet on these molecular systems. To our knowledge, the only results concern the bond length and 
the dissociation energy of Fr$_2$ which have been determined by Lim {\it et al} \cite{lim2005}
using various approaches based on relativistic coupled-cluster method or functional density theory. 
Also the van der Waals interactions between Fr, Rb, and Cs atoms have been 
computed by Derevianko {\it et al} \cite{derevianko2001}.

The purpose of this paper is to calculate the electronic properties of Fr$_2$, RbFr, and
CsFr and of their cations, including potential curves, permanent and transition
dipole moments. Our aim is also to  predict the rates for the photoassociation and for 
the formation of cold molecules, as we did previously  for all mixed alkali dimers
involving Li to Cs atoms \cite{azizi2004}.
Such predictions should be useful for upcoming experimental investigations.
Our theoretical method relies on effective potentials for modelling the Fr$^+$ ionic 
core, as in our previous paper (hereafter identified as paper I) \cite{aymar2005} 
devoted to the accurate determination of
permanent dipole moments of heteronuclear alkali dimers.

The combination of large-core pseudopotentials (namely one active valence electron per alkali atom),
 complemented by core-polarization pseudopotentials (CPP) following the idea of 
M\"{u}ller and Meyer \cite{muller1983} has proved up to now to be one of the most successful approaches
to accurately deal with the electronic structure of alkali pairs, even up to
 cesium \cite{pavolini1989,poteau1995,allouche2000,rousseau2000,korek2000,korek2000a}.
One of the advantage of the large-core+CPP approach is that it
allows full configuration interaction (CI) even in large basis sets for the valence electrons (reduced to 2 in
the case of alkali diatomics), and that the core polarization mimics the effect of 
coupled core-valence double excitations
that otherwise should be calculated in the CI. Moreover, even for compounds including cesium,
 most calculations were performed within the framework of the so-called scalar (i.e. spin-orbit averaged)
 pseudopotentials, or averaged relativistic pseudopotentials (AREP) \cite{ermler1981,teichteil1983,hurley1986}, which
take into account scalar relativistic effects (mass correction, Darwin term) \cite{cowan1976}.
The definition of quasi-relativistic pseudopotentials
 is actually  closely linked to the way electrostatic and spin-orbit coupling are treated. AREP's are very convenient for
molecular calculations, as configuration interaction is achieved in a first step,  whereas spin-orbit coupling 
responsible for multiplicity splitting is accounted for a-posteriori.
Other schemes
have been implemented in the case of heavy elements where spin-orbit coupling is large, namely 
two-component schemes \cite{christiansen1983} in the basis of spinors which treat electrostatic and spin-orbit coupling
on an equal footing in the configuration interaction.
Four-component CI calculations are also sometimes feasible \cite{pykko1986} but still require a heavy computational 
effort with respect to the
other methods. Detailed and recent overviews of the various techniques can be found in
 references \cite{balasubramian1997,hess2000,dolg2002,pykko1986,pykko1993,pykko2000,schwerdfeger2002,teichteil2004}.
In alkali diatomics, spin-orbit coupling does not directly concern the ground state but
only the excited states, and it therefore remains relatively small, even for lowest excited states of the heavy elements: 
the Cs($6p$) fine structure splitting is 554 cm$^{-1}$, and the one of Fr($7p$) is 1686 cm$^{-1}$ \cite{arnold1990},
 to be compared for instance with the one of iodine,  close to 1 eV. Core-valence correlation is a larger effect in heavy alkalis, 
especially for the lowest valence states, closer to the core. Even  more important is the
two-electron valence correlation in alkali diatomics which must be treated accurately. This is why  quasi-relativistic 
calculations with  averaged relativistic large-core pseudopotentials complemented by CPP's  proved to be very accurate for 
complexes of alkali atoms at least  up to cesium \cite{foucrault1992}.

Given this ability, it is appealing to extend this simple scheme to systems including
the heaviest atom of the series, namely Francium.
We have therefore determined a spin-orbit-averaged semi-local pseudopotential for 
francium in the large-core approach, namely with a single external electron,  using 
the shape-consistent norm-conserving technique of Durand and Barthelat
\cite{durand1974,durand1975}, complemented with core-polarization  pseudopotentials (CPP). 
We describe in detail in section \ref{sec:atom} the chosen model, that we compare to the model 
potential approach of Klapisch \cite{klapisch1969}. We
recall in section \ref{sec:mole} the main steps of the molecular calculations, and our 
results for molecular ions and neutral diatomics of Francium coumpounds. Atomic units 
will be used throughout the text, except otherwise stated.

\section{Atomic structure calculations and pseudopotential determination}
\label{sec:atom}

In 1939, M. Perey  of the Curie Institute in Paris discovered a new radioactive element 
that she named Francium \cite{perey1937}. The optical spectrum of Francium
remained unknown  until 1978 when  the D$_2$ line of  francium  was detected  by 
Liberman {\it et al} at CERN on the on-line mass separator ISOLDE
 \cite{liberman1978}. 
Then  the precise spectroscopy of francium isotopes was performed by
the same group still at CERN \cite{arnold1989,arnold1990}, while the properties of Fr
highly excited states were explored at Moscow University \cite{andreev1987}.
After 1995, spectroscopic studies were carried on with cold trapped Francium atoms 
\cite{simsarian1996a,lu1997,simsarian1998,sprouse1998,sprouse2002}.
Hyperfine splittings \cite{arnold1990,grossman1999} and radiative lifetime measurements  
\cite{simsarian1998,grossman2000} became then available. Among the energy levels, $s$
levels and Rydberg $d$ series have been the most investigated using two-step or 
two-photon excitation from the $7p$ manifold
\cite{andreev1987,arnold1989,arnold1990,simsarian1996}, as well
as the $7p_j$ and $8p_j$ multiplets. An ionization potential of 32848.972(9)cm$^{-1}$ has 
been determined from the quantum defects of $ns$ and $nd$ Rydberg levels \cite{arnold1990}. 
However the lowest $6d~_{3/2,5/2}$ levels have not been observed yet,
even if their location has been predicted using quantum defects fits of $nd_j$ 
energies \cite{grossman2000a}. Levels of higher symmetry ($f$, $g$) as well as $p$ levels
beyond $n=7,8$ remain unknown.

On the theoretical side, Dzuba {\it et al} \cite{dzuba1983,dzuba1995} used relativistic 
Hartree-Fock and many-body perturbation theory to calculate the
transition energies, fine structure intervals,
electric transition amplitudes, and addressed the issue of parity violation in this atom.
Eliav {\it et al} \cite{eliav1994} calculated ground and excited transition
energies as well as ionization potentials and electron affinities \cite{landau2001,eliav2005} for all alkali atoms
including Fr, with the relativistic coupled-cluster method.
Relativistic many-body calculations of  transition energies, hyperfine
 constants, electric-dipole matrix elements and static polarizabilites
were performed by Safronava {\it et al} for all alkali atoms
 \cite{safronova1999}. Derevianko {\it et al} \cite{derevianko2000} used
many-body perturbation theory (MBPT) and model-potential calculations
to investigate low-energy photoionization parameters for Fr.
The same model-potential approach has been adopted by Marinescu
\cite{marinescu1994a} to compute long-range parameters for all alkali pairs
but  Fr and by Marinescu {\it et al} \cite{marinescu1998} for
  a Francium atom pair. More recently,
 long-range parameters for homonuclear and heteronuclear alkali dimers including Fr
have been calculated by  Derevianko {\it et al} \cite{derevianko1999,derevianko2001}
using relativistic coupled-cluster method.

All these data represent a rich starting point for our molecular calculations which 
first impliy the determination of an accurate effective potential for the
representation of the Francium ionic core. We performed  three different atomic calculations. 
We first implemented for Francium
the parametric energy-adjusted model potential (MP) proposed by Klapisch \cite{klapisch1969},
further improved to involve $\ell$-dependent parameters (with $\ell$ the orbital moment of the 
valence electron), core polarization effects, and spin-orbit terms \cite{aymar1996}. Accurate atomic
energy levels are obtained, allowing the use of MP predicted levels to check the quality of the 
pseudopotential described below. However, the Klapisch type model potentials conserve
the nodal structure of the wavefunctions even within the core region, which is not very convenient for
molecular calculations, especially in such heavy elements.
In line with previous works on alkali diatomics, we preferred
to use a pseudopotential approach which yields nodeless atomic wave functions in the core region. 
We thus adopted a mean-field approach extraction. We first
 performed relativistic Dirac-Fock SCF calculations (DF) with 
Desclaux's package \cite{desclaux1975},  to obtain single electron energies and 
spin-orbitals considered as reference data for subsequent pseudopotential 
determination. 
We then determined a spin-orbit-averaged semi-local 
pseudopotential (PP) for Francium within the large-core approach, complemented
with core-polarization pseudopotential (CPP) which will be used in molecular
calculations of section \ref{sec:mole}.

\subsection{Parametric model potential (MP)}

Such a model is known to be efficient for instance to calculate multiphoton cross 
sections of heavy alkali atoms K and Cs \cite{aymar1981,aymar1982} and complicated
spectra of heavy  alkaline-earth atoms \cite{aymar1996}. A similar model
potential without spin-orbit has been also used by other authors for the determination of 
the van der Waals interaction between alkali atoms \cite{marinescu1994},
or of photoionization parameters in Fr \cite{derevianko2000}.
The semi-local model potentiel accounting for the interaction between the ionic core and 
the external electron depends on the empirical parameters $\alpha_{i}^{\ell}$ and 
contains a polarization term involving the static dipole polarizability (core-
polarization contribution) of the ionic core $\alpha_d$:     

\begin{equation}
V_{\ell}(r)=  -\frac{1}{r}\lbrace 1+(Z-1) \exp{(-\alpha_1^{\ell} r)} + 
\alpha_2^{\ell}r \exp{(-\alpha_3^{\ell} r)} \rbrace- \frac{\alpha_{d}}{2r^4}\lbrace 1- \exp{
\lbrack - (r/r_c^{\ell} )^6 \rbrack} \rbrace
\label{eq:klap}
\end{equation} 

where $Z$ is the nuclear charge, and $r_c^{\ell}$ are cut-off radii truncating the polarization 
potential at short electronic distances $r$. We use the theoretical value for $\alpha_d$=20.38~au
\cite{lim2002} obtained with fully relativistic coupled cluster method, in very good agreement 
with the value $\alpha_d$=20.41~au computed using relativistic random phase approximation \cite{derevianko1999}.
 The one-electron radial function 
$P_{n\ell j}$ (where $j$ is the total angular momentum of the external electron) is the solution 
of the radial Schr\"{o}dinger equation with eigenvalue $\epsilon_{n \ell j}$:
                                                                 
\begin{equation}
\lbrack -\frac{1}{2}\frac{d^2}{dr^2}+ \frac{\ell(\ell+1)}{2r^2}+V_{\ell}(r)+V_{so}^{s\ell j}(r)-
\epsilon_{n \ell j} \rbrack  P_{n \ell j}= 0
\label{eq:pnj}
\end{equation} 

The spin-orbit scalar operator $V_{so}^{s\ell j}(r)$ is explicitely included and is approximated by:

\begin{equation}
V_{so}^{s\ell j}(r) = \frac{\alpha^2}{2} \vec{s}.\vec{\ell}\frac{1}{r}\frac{dV}{dr} {\biggr (}
1-\frac{\alpha^2}{2} V(r){\biggr )}^{-2}
\label{eq:vso}
\end{equation}

where $\alpha$ is the fine structure constant. The last factor    
suggested by the Dirac equation \cite{condon1935} is included, to ensure that the solutions of 
the radial Schr\"{o}dinger equation are well-defined when $r \rightarrow 0$.
The empirical parameters $\alpha_{i}^{\ell}$ and r$_{c}^{\ell}$ are adjusted until the
energies $\epsilon_{n \ell j}$  agree with the experimental atomic energies (table \ref{tab:model}).

\begin{table}[h]
\center
\begin{tabular}{|c|c|c|c|c|} \hline
$\ell$& $\alpha_1^{\ell}$& $\alpha_2^{\ell}$& $\alpha_3^{\ell}$& $r_c^{\ell}$ \\ \hline
0& 4.29673059&13.58990866&1.54459081&1.35534853\\
1& 4.10987255&5.87755911&1.36939901&1.22513294\\
2&3.39019850&0.78543635&0.92256617&1.96472431\\ \hline
\end{tabular}
\caption{Empirical parameters  involved in the model potential
of eq.\ref{eq:klap} describing the Francium ionic core. The value $\alpha_d$ =20.38~a.u. 
\cite {lim2002} is introduced in the polarization potential.}
\label{tab:model}
\end{table}

Atomic binding energies of Francium obtained by the MP approach are
compared with experimental values and with several other
computations in table \ref{tab:energy}: the
MPBT calculations of Dzuba {\it et al} performed whithin the Bruckner expansion 
\cite{dzuba1995},
the relativistic coupled-cluster (CC) calculations of Eliav {\it et al}
\cite{eliav1994}, the relativistic many-body calculations of Safronova {\it et al}
\cite{safronova1999}, 
and the results of the Quantum Defect Theory (QDT) combined with MBPT and model potential calculations of 
Derevianko {\it et al} \cite{derevianko2000}.

The present MP energies  are found in good agreement with the available observed energies, and 
are generally more accurate than the other theoretical predictions. The fine structure 
splitting of the $6d$ level, $\Delta E_{so}=272.1$~cm$^{-1}$, is 
found larger than in the CC value of ref.\cite{eliav1994}, $\Delta E_{so}=200$~cm$^{-1}$, and also disagrees 
 with the values predicted using quantum defects fits of $nd_J$ 
energies \cite{grossman2000a} ($\Delta E_{so}=188$~cm$^{-1}$). Our predictions for all the other unobserved
$np_j$ levels agree with other theoretical values, giving in particular similar fine-structure
splitting  for $8p$ to $10p$ levels. These predictions for energy levels will be used in the next paragraph.

\begin{table}
\center
{\small
\begin{tabular}  {|c|c|c|c|c|c|c|c|c|c|c|c|c|c|}
\hline
Level&Exp&MP&$\Delta E$&\cite{dzuba1995}&$\Delta E$&\cite{eliav1994}&$\Delta 
E$&\cite{safronova1999}&$\Delta E$&\cite{derevianko2000}(1)&\cite{derevianko2000}(2) \\
&&Present work&&&&&&&&&\\ \hline
7s &-32848.87&-32848.87&0.05&-32762& 86&-32839&9.8&-32735&113&           &\\
7p-&-20611.46&-20613.03&-1.5&-20654&-42&-20584& 27&-20583& 28&           &\\
6d-&         &-16684.56&    &-16623&   &-16370&   &      &   &           &\\
6d+&         &-16412.47&    &-16423&   &-16194&   &      &   &           &\\
7p+&-18924.87&-18922.33& 2.5&-18926&-1.2&-18913&11&-18907& 17&           &\\
8s &-13116.62&-13116.39& 0.2&-13082& 35& -13121&-4&-13051& 65&           &\\
8p-&9735.91  & -9730.38&5.55& -9742&-6& -9719& 17& -9712&  23&        &\\
8p+ &9190.56.0 & -9199.62&-9.0& -9188&-29 &  -9158& 32& -9176& 14&   &\\
7d-& -8604.04& -8605.25&-1.2& -8663&-58&       &  &      &   &           &\\
7d+& -8515.57& -8514.55& 1.0& -8574&-58&       &  &      &   &           &\\
9s & -7177.85& -7177.94&-0.09&-7160& 18&       &  & -7148& 29&           &\\
9p-&         & -5733.61&    & -5736&   &       &  & -5724&   &-5748& -5737\\
9p+&         & -5493.70&    & -5485&   &       &  & -5480&   &-5496 &-5487\\
8d-& -5248.22& -5246.58& 1.6& -5261&-12&       &  &      &   &           &\\
8d+& -5203.50& -5204.42&-0.9& -5218&-15&       &  &      &   &           &\\
10p-&        & -3787.81&    & &   &       &  & -3782&   &-3800& -3790\\
10p+&        & -3658.66&    & &   &       &  & -3650&   &-3662 &-3655\\
10s &-4538.26& -4538.29&-0.03&-4534&  4&       &  & -4522& 16& &          \\ \hline
\end{tabular}
}
\caption{Theoretical binding energies of Francium  (in cm$^{-1}$) obtained with the present 
MP approach, compared with other relativistic calculations and experimental energies from refs.
 \cite {arnold1990,grossman2000a}. Differences $\Delta E$ between calculated and experimental energies 
are also displayed. Levels labelled with a $\pm$ sign means $j=\ell \pm 1/2$.}
\label{tab:energy}
\end{table}

\subsection{Pseudopotential approach (PP)}

We actually used this approach to perform the molecular calculations 
for Fr$_2$ and RbFr and CsFr compounds. The ionic core is described by a semi-local 
$\ell$-dependent pseudopotentiel (PP)
 \cite{durand1974,durand1975} employed in paper I:

\begin{equation}
W_\ell(r) =-1/r+\sum_{i} C_{i, \ell} r^{n_{i, \ell}} \exp(-\alpha_{i, \ell}r^2)
\label{eq:vps}
\end{equation}

 As explained in paper I, and in contrast with the MP approach, the parameters 
$C_{i \ell}$ and $\alpha_{i \ell}$ in equation 
 \ref{eq:vps} for Rb and Cs ionic cores are adjusted to reproduce the energies and valence 
 orbitals of an all-electron Hartree-Fock  Self-Consistent Field (SCF) calculation for the atomic ground
 state.

The Francium PP parameters (table \ref{tab:pseudo}) are chosen in order to match 
the results of the all-electron relativistic Dirac-Fock (DF) SCF calculations 
performed with Desclaux's package \cite{desclaux1975}.
More precisely, we determine the large components of the
Dirac-Fock solutions for the valence levels $7s_{1/2}$, $7p_{1/2,3/2}$,
$6d_{3/2,5/2}$ and $5f_{5/2,7/2}$, and we adjust the PP parameters
 to match the corresponding {\it spin-orbit-averaged} orbitals.
Relativistic corrections to atomic radial wave functions
such  mass-velocity, and the Darwin terms are automatically introduced in
Dirac-Fock orbitals and thus in the pseudopotential
 \cite{cowan1976}, which keeps the simple expression of equation \ref{eq:vps}.
As discussed by several authors on Cs or Fr exemples
\cite{dzuba1983,dzuba1995,safronova1999}, zero-order
Dirac-Fock energies do not include correlation effects
necessary to reproduce the atomic energy level structure, taken in account
through MPBT in ref.\cite{dzuba1983,dzuba1995,safronova1999,eliav1994}. However, while
intra-core correlation effects are hardly accessible via CI, inter-shell core-valence
correlation effects can be described via the core-polarization operator.
Let us note finally that an {\it ab initio}  small-core pseudopotential (involving an ionic core with 78 
electrons, instead of 86 in the present work) including
relativistic effects has been obtained  by Ermler \cite{ermler1991}, 
 and further used by Lee {\it et al} to compute ground state potential
energy curves of MO oxydes with (M=Rb,Cs,Fr) \cite{lee2001}. However the authors have not extended their investigations
towards francium-alkali pairs.

\begin{table}
\center
\begin{tabular}  {|c|c|c|c|}
\hline
$\ell$ & $\alpha_{i, \ell}$&  $c_{i, \ell}$& $n_{i, \ell}$\\  \hline
0& 0.784936& 14.194790&1\\
1&0.170429&0.352240&0\\
& 0.170429& 0.309530& 1\\
2&0.258663&-3.442104&0\\
&0.258663& 4.652569&-1\\ \hline
\end{tabular}
\caption{Parameters of the pseudopotential of eq.\ref{eq:vps} representing the Francium 
core, implemented in the present work.}
\label{tab:pseudo}
\end{table}

The Dirac-Fock orbitals for the fine structure doublets $7p_{1/2,3/2}$,
$6d_{3/2,5/2}$, and $5f_{5/2,7/2}$ are plotted in figure \ref{fig:dforb}. It
is clear that even if differences between $7p_j$ orbitals are visible, this figure
suggests that spin-orbit averaged orbitals can be used to set up a pseudopotential for 
the Francium ionic core with averaged relativistic effects. Such orbitals are drawn in 
figure \ref{fig:psorb} for the $7s$, $7p$, and $6d$ levels, as computed via
the present MP, DF, and PP approaches. As expected, 
the mean-field PP pseudo-orbitals perfectly match the DF spin-orbit 
averaged orbitals in the region outside 
the core, i.e. the outer lobe of the wave function is well reproduced. In contrast, the MP orbitals
are shifted inwards with respect to the DF orbitals, except for the 7s wavefunction.
The same trend (not shown here) is actually observed with the nodeless PP pseudo-orbitals when the core
polarization effects (see next section for the modelling of these effects) are included, compared to those
pseudo-orbitals without core polarization. As the MP orbitals intrinsically contain core polarization effects,
we suspect that this explains their observed pattern.

\begin{figure}
\center
\caption{Dirac-Fock orbitals for the (a) 7p$_{1/2,3/2}$, (b) 6d$_{3/2,5/2}$, (c) 5f$_{5/2,7/2}$ 
levels of Francium. The orbital with the lowest (largest) $j$ value is drawn with full (dotted) 
line.}
\label{fig:dforb}
\end{figure}

\begin{figure}
\center
\caption{Spin averaged orbitals for the (a) $7s$, (b) $7p$, (c) $6d$ levels of Francium, calculated 
with various approaches: Dirac-Fock (dashed line), present pseudopotential (red full line), present 
model potential (blue dotted line).}
\label{fig:psorb}
\end{figure}

\subsection{PP results for atoms}

The external electron in each atom (Rb, Cs, or Fr) is described 
with a Gaussian basis set. In order to check the accuracy against the size of the basis 
set, we proceed as in paper I, defining two basis sets for Rb and Cs: we first use a 
[$7s4p5d$] contracted to a [$6s4p4d$] basis, identical to the one introduced by Pavolini 
{\it et al} \cite {pavolini1989}  previously refered to as basis "A" in paper I.
Then we consider for Rb the large [$9s6p6d4f$] basis of uncontracted Gaussian functions resulting 
in 138 molecular orbitals (basis "B" in paper I), and we elaborate a similar
[$9s6p6d4f$] basis for Cs refered to as "B'" in the following (table \ref{tab:basis}).
For the Fr atom, we use a large uncontracted [$9s9p9d$] Gaussian basis set yielding 162 
molecular orbitals (table \ref{tab:basis}). Molecular calculations for RbFr and CsFr then 
combine the Fr basis with one of the above basis sets for Rb and Cs. 

\begin{table}[h]
\center
\begin{tabular} {|c|c|c|l|}
\hline
Atom&Basis&$\ell$&Exponents \\ \hline
Cs&[$9s6p6d4f$] &s&2.35,1.492561,0.824992,0.5,0.234682,0.032072,0.013962,0.005750,0.0025 \\
  &           &p& 0.22,0.128,0.040097, 0.03,0.014261,0.004850\\
  &           &d& 0.29,0.12,0.096036,0.026807,0.009551,0.004\\ 
  &           &f& 0.2,0.1,0.05,0.005\\ \hline
Fr&[$9s9p9d$] &s& 1.8,0.480468,0.369521,0.2,0.11230,0.053409,0.018240,0.006,0.002\\
  &           &p& 0.22,0.120,0.0655,0.03,0.0162,0.008,0.00443,0.002,0.001\\
  &           &d& 1.3,0.6,0.30,0.196894,0.067471,0.027948,0.010712,0.00300,0.001\\ \hline
\end{tabular}
\caption{Exponents of the Gaussian functions introduced in the [$9s6p6d4f$] basis for Cesium (basis "B'"), 
and the [$9s9p9d$] basis for Francium.}
\label{tab:basis}
\end{table}

Core-valence correlation is treated by adding a phenomenological core polarization 
operator $W_{CPP}$ to the Hamiltonian of the valence electrons \cite{muller1983}
in the field of the $M^+$ and $M'^+$ ionic cores, and : 

\begin{equation}
W_{CPP}=-\frac{1}{2}\alpha_d^{M^+}\vec{f_M}^2-\frac{1}{2}\alpha_d^{M'^+}\vec{f_M'}^2
\end{equation}

where $\alpha_d^{M^+}$ is the static dipole polarizability of the M$^+$ ion, and 
$\vec{f}_M$ (resp. $\vec{f}_M'$) is the total electric field seen by $M^+$ (resp.  
$M'^+$) due to the two electrons {\it i} and the partner ion $M'^+$ (resp.  $M^+$):

\begin{equation}
\vec{f}_M=\sum_{i=1,2}
-\frac{\vec{r}_{iM}}{r_{iM}^3}\Theta(r_{iM})+\frac{Z_{M'}\vec{R}_{M'M}}{R_{M'M}^3}
\end{equation}

Experimental values for $\alpha_d^{M^+}$ are available for Cs and Rb as 
reported in paper I. We introduce a dependence on the electronic orbital momentum 
$\ell$ in the effective 
core polarization potential \cite{foucrault1992}, through the cut-off functions 
$\Theta_\ell$ to prevent the CPP matrix elements from diverging at the origin. As in ref. 
\cite {foucrault1992}, $\Theta_l$ is taken as a step function vanishing below the
$\ell$-dependent cut-off radii $\rho_{\ell}$.

For each basis set and a fixed value of $\alpha_d^{M^+}$, the cut-off radii are
adjusted to reproduce the experimental energies \cite{moore1949,moore1952,moore1958} 
of the lowest $s$, $p$ and $d$ levels of Cs and Rb (Table \ref{tab:cutoff}).
As no experimental determination for the Fr$^+$ polarizability is available, we use like
in the previous MP approach $\alpha_d^{Fr^+}=$20.38~a.u. \cite{lim2002}. Another estimate 
$\alpha_d^{Fr^+}=$23.2~a.u. was obtained from an extrapolation of known
core polarizabilities of the lighter alkali atoms \cite{derevianko2000}, and we will 
discuss in the next section the influence of this parameter. As for Rb and Cs, the 
cut-off radii are adjusted on the experimental atomic energies 
\cite{arnold1990,grossman2000a}, for each above value of $\alpha_d^{Fr^+}$ (Table
\ref{tab:cutoff}). The sensitivity of the results with the cut-off radii values is such 
that the influence of the polarizability can be balanced to provide a good representation 
of energy levels with both sets A and B. 

\begin{table}
\center
\begin{tabular} {|c|c|c|c|c|c|} \hline
Atom&Series& $\alpha_d^{M^+}$&$\rho_s$&$\rho_p$&$\rho_d$\\ \hline
Fr&A&20.38 \cite{lim2002} & 3.16372&3.045&3.1343 \\ \hline
  &B&23.2\cite{derevianko2000}& 3.343&3.435&3.292 \\ \hline
Rb&A&9.245&2.5213&2.279&2.511\\ \hline
  &B&9.245&2.5538&2.3498&2.5098 \\ \hline
Cs&A&15.117&2.6915&1.8505&2.807 \\
  &B'&16.33&2.0081&2.6865&2.83518 \\
\hline
\end{tabular}
\caption{Dipole polarizabilities $\alpha_d^{M^+}$ and  cutoff parameters 
$\rho_{\ell}$ introduced in the effective core polarization potential (CPP) for M$^+=$Fr, 
Rb, Cs. Values for Rb (basis A and B) and Cs (basis A) are recalled from paper I, 
while new values for the Cs B' basis are reported.}
\label{tab:cutoff}
\end {table}

\begin{table}[t]
\center
\begin{tabular} {|c|c|c|c|c|c|c|c|c|}\hline
Level& Exp& with $\alpha_d^{Fr^+}=$20.38~a.u.& $\Delta$E&with $\alpha_d^{Fr^+}=$23.2~a.u. &$\Delta E$\\ \hline
$7s$  & -32848.87&  -32848.87&  0.002& -32848.87&     0.002 \\
$7p$   &-19487.07&  -19486.93& 0.143& -19487.18&    -0.110\\
$6d$   & -16471.07&  -16470.92&  0.147& -16470.94&     0.128\\
$8s$   & -13116.62& -13161.02&  -44.& -13136.97&   -20.\\
$8p$   &  -9372.35&   -9448.16&   -76.& -9424.96&   -53.\\
$7d$   &  -8550.96&   -8394.77&   156.& -8371.59  & 179.\\
$9s$  &   -7177.85&   -7161.96&    16.& -7151.28 &   26. \\
$9p$   &  -5573.18 &  -5619.67&   -46.& -5605.27&   -32.\\
$8d$   &    -5221.39&  4981.02& 240&4963.01&258\\ \hline
\end{tabular}
\caption {Computed binding energies of atomic Francium atom in cm$^{-1}$
with the two available Fr$^+$ polarizabilities, compared to experimental values
\cite {arnold1990,grossman2000a}, and corresponding energy differences
$\Delta$E. For the unobserved $9p$ level our reference is the spin-averaged 
value predicted by the model-potential approach described earlier in the text.} 
\label{tab:binding}
\end {table}

\section {Molecular calculations}
\label{sec:mole}

As in paper I, we used an automatized procedure based on the CIPSI package \cite{cipsi}.
 The molecular  orbitals are determined by restricted single electron calculations 
including the CPP \cite{foucrault1992}, namely corresponding to [XY]$^+$ systems, providing
the potential curves for the relevant molecular cations. A full valence configuration 
interaction (CI) is then performed for each involved molecular symmetry, providing 
potential curves and permanent and transition dipole moments.
In the case of neutral pairs,  two-electron full CI is achieved in the framework of
a single-component scheme. Clearly, further investigations using two- and
four component schemes or larger cores would be interesting, although not easily
tractable even at present, especially in the case of excited states, if one wants to 
include excitations corresponding to core polarization effects. Such calculations would 
anyway imply the use of two- or four-component pseudopotentials, which was not achieved 
here.

In addition to the explicit above calculations, it is also important to estimate the 
magnitude of the deviation from the Coulomb repulsion of core-core interactions, which
could be significant for the heavier and larger alkali atoms. The most significant effect is 
the repulsive overlap of the cores, which can be either extrapolated from exponential
formulas valid for the lighter species \cite{pavolini1989}, or derived from a calculation
of [XY]$^{++}$ at the Hartree-Fock level, for instance within the frozen core approximation
\cite{rousseau2000,korek2000,jeung1997} which is adapted here as the core polarization
is accounted for through an effective potential. The main influence of this term is to make
the short-range repulsive part of the potentials steeper, while it dies out in the
region of the minimum of the ground state. According to ref.\cite{jeung1997}, this term
contributes for about 15 cm$^{-1}$ at the equilibrium distance of the ground
state of Na$_2$, K$_2$, Rb$_2$ and Cs$_2$, and slightly more ($\approx$40 cm$^{-1}$) for RbCs.
A comparable estimation is probably appropriate for the Francium compounds studied here.

The contribution of the core-core dispersion energy can be estimated
via the London formula (see for instance ref.\cite{pavolini1989}):

\begin{equation}
V_{cc}^{disp}(R)=-\frac{3\alpha_d^{X^+} \alpha_d^{Y^+}}{2R^6} \frac{E_I^{X^+} E_I^{Y^+}}{E_I^{X^+}+E_I^{Y^+}}
\label{eq:disp}
\end{equation}

where $E_I^{X^+}$ and $E_I^{Y^+}$ are the ionisation energies of the X$^+$ and Y$^+$ 
ions respectively. At the equilibrium distance of the ground state, this term deepens 
the potential by about 27, 16 and 18 cm$^{-1}$ for Fr$_2$, RbFr and CsFr respectively.
It basically compensates the core-core term above.

\subsection{Potential curves of Fr$_2^+$, RbFr$^+$, CsFr$^+$}
\label{ssec:cations}

We display in figure \ref{fig:fr2+} the potentiel curves of the ground state (1)$^2
\Sigma_g^+$ and the lowest (1)$^2\Sigma_u^+$ state of Fr$_2^+$. We see in figure
\ref{fig:fr2+}a that the choice of the large Fr$^+$ polarizability
$\alpha_d^{Fr^+}=$23.2~a.u. induces a potential well deeper by about 16~cm$^{-1}$ than 
with $\alpha_d^{Fr^+}=$20.38~a.u. . Its effect is then rather small, possibly within the 
usual inaccuracy which was previously observed for this type of quantum chemistry calculations.

Since no experimental result is presently available, we qualitatively compare our results  for 
the Fr$_2^+$ ground state with those of
Rb$_2^+$ and Cs$_2^+$ in the main panel of Figure \ref{fig:fr2+}, and in Table 
\ref{tab:ion_wells}. 
The well depth of the ground state indeed decreases for increasing mass of the atom, as 
it can be observed over the whole sequence of homonuclear alkali ions. In contrast the Fr$_2^+$ 
equilibrium distance is found slightly smaller the Cs$_2^+$ one, and  larger than that of
Rb$_2^+$. Note that a similar trend was also found by Lim {\it et al} \cite{lim2005}, but the large dispersion
of the dissociation energy found by these authors does not allow an easy comparison with our results.
These results illustrate that the properties of the Fr atom are not reducible to
extrapolations from the lightest ones due to relativistic effects. This is
fingerprinted by the size evolution of the lowest $ns$ valence orbital,
the $7s$ orbital of Francium being more contracted than the $6s$ one of cesium,
whenever relativistic effects are taken into account. Indeed, Dirac-Fock calculations
yields averaged radii values of 5.91~a$_0$ for the $7s$ orbital in
Francium and 6.08~a$_0$  for the $6s$ orbital in  cesium, while the non-relativistic HF
calculation reverses this hierarchy yielding 6.25~a$_0$ for Francium $7s$, and 5.85~a$_0$ 
for Cesium 6s. Such orbital contractions were illustrated on atoms
 by Desclaux \cite{desclaux1973}.
In the case of alkali's, these contractions are further  correlated with
an increase of the  atomic ionization potentials as discussed later on in the paper. This
behaviour illustrates on  francium compounds the influence of relativistic effects
on the chemical bonding whenever heavy atoms are involved.

A similar pattern is observed for the
(1)$^2\Sigma_u^+$ state (\ref{fig:fr2+}b). It is interesting to note that such a shallow
long-range potential well is predicted for all ionic alkali dimers including
 heteronuclear ones \cite{aymar2003,azizi2006,magnier1996a}, and 
could be used as an alternative route for the detection through resonant ionization 
of ultracold molecules \cite{aymar2005,dion2002}.

\begin{figure}
\center
\vspace{1cm}
\caption{Main panel: potential curves of the ground state $1^2\Sigma_g^+$ (full lines) and the lowest 
$1^2\Sigma_u^+$ state (dashed lines) of Fr$_2^+$ (black lines), compared with those of Rb$_2^+$ 
(red lines with open circles) and Cs$_2^+$ (blue lines with closed circles).
All curves are referenced to the same dissociation energy taken as the origin.
(a) Comparison of calculations with $\alpha_d^{Fr^+}=$20.38~a.u. \cite{lim2002}
(full line) and with $\alpha_d^{Fr^+}=$23.2~a.u. \cite{derevianko2000} 
(dot-dashed line). (b) Blow up of the long-range well predicted in the $1^2\Sigma_u^+$. 
The main and (b) panels display calculations with $\alpha_d^{Fr^+}=$20.38~a.u., and with 
basis B and B' for Rb$_2^+$ and Cs$_2^+$ repectively.}
\label{fig:fr2+}
\end{figure}

\begin{table}[t]
\center
\begin{tabular} {|c|c|c|c|c|}\hline
$(1)^2\Sigma_g^+$&Basis&$D_e$ (cm$^{-1}$)&$R_e$ ($a_0$) \\\hline
Fr$_2^+$&(a)&5537&9.61  \\
        &(b)&5552&9.65  \\\hline
Rb$_2^+$&A&6138&9.08  \\
        &B&6208&9.05  \\\hline
Cs$_2^+$&A&5950&9.75  \\
        &B'&5977&9.85  \\\hline \hline
$((1)^2\Sigma_u^+$&Basis&$D_e$ (cm$^{-1}$)&$R_e$ ($a_0$) \\\hline
Fr$_2^+$&(a)&72.3&24.0  \\
        &(b)&72.6&24.0  \\\hline
Rb$_2^+$&A&81.7&23.0  \\
        &B&82.1&23.0  \\\hline
Cs$_2^+$&A&82.7&24.3  \\
        &B'&84.6&24.3  \\\hline \hline
$(1)^2\Sigma^+$&Basis&$D_e$ (cm$^{-1}$)&$R_e$ ($a_0$) \\\hline
RbFr$^+$&A&5439&9.3  \\
        &B&5450&9.3  \\\hline
CsFr$^+$&A&5054&9.68  \\
        &B'&5052&9.72  \\\hline
RbCs$^+$&A,A&5053&9.40  \\
        &B,B'&5090&9.41  \\\hline \hline
$((2)^2\Sigma^+$&Basis&$D_e$ (cm$^{-1}$)&$R_e$ ($a_0$) \\\hline
RbFr$^+$&A&205.9&19.65  \\
        &B&206.1&19.65  \\\hline
CsFr$^+$&A&291.3&19.07  \\
        &B'&295.7&19.06  \\\hline
RbCs$^+$&A,A&381.6&17.83  \\
        &B,B'&389.5&17.84  \\\hline \hline
\end{tabular}
\caption {Well depth $D_e$ and equilibrium distance $R_e$ ($a_0$=0.0529177~nm)
of the potential wells obtained in the present work for the two
lowest $^2\Sigma$ states of Fr ionic compounds. Results for Rb$_2^+$, Cs$_2^+$ and 
RbCs$^+$ are also displayed for comparison. (a) and (b) results for Fr$_2^+$
correspond to $\alpha_d^{Fr^+}=$20.38~a.u. \cite{lim2002} and with 
$\alpha_d^{Fr^+}=$23.2~a.u. \cite{derevianko2000} respectively. Results for heteronuclear
systems are obtained with
$\alpha_d^{Fr^+}=$20.38~a.u.. Note that results with basis B' for Cs$_2^+$ and
(B,B') basis results for RbCs$^+$ are new, while those Rb$_2^+$ were already
given in ref.\cite{azizi2004}.}
\label{tab:ion_wells}
\end {table}

We performed the same analysis for the two lowest states of RbFr$^+$ and CsFr$^+$
compared with those of RbCs$^+$ (figure \ref{fig:mfr+}and Table \ref{tab:ion_wells}). 
Among the three atoms, Rb has 
the largest ionization potential and Cs the lowest one. The ground state $(1)^2\Sigma^+$
of RbFr$^+$ (resp. CsFr$^+$) is correlated to Rb(5s)+Fr$^+$ (resp. Fr(7s)+Cs$^+$) and the
$(2)^2\Sigma^+$ to Fr(7s)+Rb$^+$ (resp. Cs(6s)+Fr$^+$). There is no clear trend for the
hierarchy among well depths of the three systems.
An analysis of the ground state of the whole ensemble of heteronuclear alkali molecular 
ionic species \cite{azizi2006} shows that the NaLi$^+$ has the deepest well (more than 
8000~cm$^{-1}$) with the smallest equilibrium distance (around 6.5$a_0$). The RbCs$^+$, 
KCs$^+$ and KRb$^+$ ions have comparable well depth (in the 5000-6000~cm$^{-1}$ range) 
and equilibrium distances (around 8.5-9$a_0$), while the remaining species exhibit well 
depth ranging between 3500 and 5000~cm$^{-1}$, with comprable equilibrium distance around 
8-9$a_0$. The RbFr$^+$ and CsFr$^+$ ions fit into the second category. As already  found
for RbCs$^+$ in ref.\cite{aymar2003}, a shallow long-range well is present in the 
(2)$^2\Sigma^+$ potential curve of RbFr$^+$ and CsFr$^+$ ions (figure \ref{fig:mfr+}b)
similar to the well of (1)$^2\Sigma_u^+$ potential of homonuclear dimers. The difference induced by the
different choices of parameters within these compounds is illustrated in figure 
\ref{fig:mfr+}c,d. As in Fr$_2^+$, the larger Francium polarizability yields a deeper
potential well. Increasing the size of the Rb basis has the same effect, as the RbFr$^+$
well depth is increased by about 11~cm$^{-1}$. In contrast, increasing in the same way
the size of the Cs basis slightly changes the CsFr$^+$ equilibrium distance, but not its depth.

\begin{figure}
\center
\caption
{Potential curves of the ground state $(1)^2\Sigma^+$ (panel a) and of the  
$(2)^2\Sigma^+$ state (panel b) of RbFr$^+$ (black lines) and CsFr$^+$ (full
red lines with open circles), compared with  those of RbCs$^+$ (blue lines with closed circles). 
The  energy origin is taken in both cases at the dissociation limits of the potentials. 
Calculations reported in panels a and b are performed with $\alpha_d^{Fr^+}=$20.38~a.u. 
\cite{lim2002}, and basis B and B' for Rb and Cs respectively. 
(c) Ground state of the RbFr$^+$ ion: calculations with $\alpha_d^{Fr^+}=$20.38~a.u.
and basis A for Rb (full line), with $\alpha_d^{Fr^+}=$23.2~a.u. \cite{derevianko2000} and
basis A (Rb) (black dot-dashed line), and with $\alpha_d^{Fr^+}=$20.38~a.u. and 
basis B (Rb) (full line with circles). (d) Same as (c) for CsFr$^+$ ion.}
\label{fig:mfr+}
\end{figure}

\subsection {Potential curves and permanent dipole moments for Fr$_2$, RbFr, CsFr lowest states}

As no experimental data are available, the neutral systems involving Fr are analyzed 
along the same lines than the molecular ions in the previous section. 
Figure \ref{fig:fr2} shows the lowest Fr$_2$ potential curves dissociating into $7s+7s$, 
compared with those of Rb$_2$ and Cs$_2$. The influence of the value of the Fr$^+$ 
polarizability is clearly larger than in the molecular ion, as two electrons  now 
contribute to the polarization of the cores (Figure \ref{fig:fr2}b,d): the depth obtained 
with $\alpha_d^{Fr^+}=$23.2~a.u. \cite{derevianko2000} is larger by about
80~cm$^{-1}$, and by about 12~cm$^{-1}$ for the
$(1)^1\Sigma_g^+$ ground state and for the $(1)^3\Sigma_u^+$ state respectively. 
As for molecular ions the the well depth increases with 
increasing mass, while and the equilibrium distance increases along the series
 Rb$_2$, Fr$_2$, Cs$_2$, again in agreement with ref.\cite{lim2005}.

The lowest potential curves of RbFr and CsFr are displayed in Figure \ref{fig:mfr},
compared with RbCs ones. In contrast with the related molecular ions, the RbCs ground 
state and lowest triplet state are found deeper than those of the Fr compounds. This suggests that
the overall energy associated to the core polarization strongly depends not only on the 
polarizability of the individual atoms, but also on the difference of their ionization
potentials, in a somewhat irregular way along the series of mixed alkali dimers.
The characteristics of these wells are reported in Table \ref{tab:wells}, including those 
for Rb$_2$, Cs$_2$, RbFr, CsFr, and RbCs. As in the previous section, Rb$_2$ results
of type B were already shown in paper I, while those for Cs$_2$ and RbCs are new. 

\begin{figure}
\center
\caption {Potential curves of the ground state $(1)^1\Sigma_g^+$
  (panel a) and the lowest $(1)^3\Sigma_u^+$ state (panel b) of Fr$_2$
  (full line with circles) calculated with
  $\alpha_d^{Fr^+}=$20.38~a.u. \cite{lim2002}, compared with those of
  Rb$_2$ with basis B (blue dot-dashed line) and Cs$_2$ with basis B'
  (red dashed line).  The energy origin is taken at the dissociation
  limits of the potentials.  (b) Ground state of the Fr$_2$ dimer
  obtained with with $\alpha_d^{Fr^+}=$20.38~a.u.  (full line), and
  with $\alpha_d^{Fr^+}=$23.2~a.u. \cite{derevianko2000} (dashed
  line). (d) Same as (b) for the $(1)^3\Sigma_u^+$ state.}
\label{fig:fr2}
\end{figure}

\begin{figure}
\center
\caption{Potential curves of the ground state $(1)^1\Sigma^+$ (panel a) and the lowest 
$(1)^3\Sigma^+$ state (panel b) of RbFr (full black line) and CsFr (dashed red line),
compared with those of RbCs (blue dots). The energy origin
is taken at the dissociation limit of the potentials. 
We used $\alpha_d^{Fr^+}=$20.38~a.u. and basis B and B' for Rb abd Cs respectively.}
\label{fig:mfr}
\end{figure}

\begin{table}[t]
\center
\begin{tabular} {|c|c|c|c|}\hline
$(1)^1\Sigma_g^+$&Basis&$D_e$ (cm$^{-1}$)&$R_e$ ($a_0$) \\\hline
Fr$_2$&(a)&3498&8.45  \\
      &(b)&3576&8.45  \\\hline
Rb$_2$&A&3861&7.9  \\
      &B&4039&7.9  \\\hline
Cs$_2$&A&3703&8.65  \\
      &B'&3787&8.69  \\\hline \hline
$(1)^3\Sigma_u^+$&Basis&$D_e$ (cm$^{-1}$)&$R_e$ ($a_0$) \\\hline
Fr$_2$&(a)&188.5&12.5  \\
      &(b)&200.8&12.5 \\\hline
Rb$_2$&A&233.6&11.6  \\
      &B&258.9&11.6  \\\hline
Cs$_2$&A&255.0&12.0  \\
      &B'&286.8&12.0  \\\hline \hline
$(1)^1\Sigma^+$&Basis&$D_e$ (cm$^{-1}$)&$R_e$ ($a_0$) \\\hline
RbFr&A&3654&8.2  \\
    &B&3690&8.2  \\\hline
CsFr&A&3553&8.55  \\
    &B'&3576&8.57 \\\hline
RbCs&A,A&3655&8.27  \\
    &B,B'&3921&8.29  \\\hline \hline
$((1)^3\Sigma^+$&Basis&$D_e$ (cm$^{-1}$)&$R_e$ ($a_0$) \\\hline
RbFr&A&207.2&12.0  \\
    &B&209.8&12.0  \\\hline
CsFr&A&209.5&12.36  \\
    &B'&217.9&12.33  \\\hline
RbCs&A,A&253.7&11.70  \\
    &B,B'&273.1&11.65  \\\hline \hline
\end{tabular}
\caption
{Well depth $D_e$ and equilibrium distance $R_e$
 of the potential wells obtained in the present work
for the ground state and the lowest triplet state of Fr neutral compounds. Results for
Rb$_2$, Cs$_2$ and RbCs are also displayed for comparison. (a) and (b) results for 
Fr$_2$ correspond to $\alpha_d^{Fr^+}=$20.38~a.u. \cite{lim2002} and with
$\alpha_d^{Fr^+}=$23.2~a.u. \cite{derevianko2000} respectively. Results for heteronuclear 
systems are obtained with $\alpha_d^{Fr^+}=$20.38~a.u.. . Note that results with basis
B' for Cs$_2$ and with (B,B') basis for RbCs are new, while those Rb$_2$ were already
given in paper I.}
\label{tab:wells}
\end {table}

The permanent dipole moment of RbFr and CsFr are given in Figure \ref{fig:permdip} for
the ground state and the lowest triplet state, respectively and compared with the results from paper I for
RbCs. Note that results are not significantly different when using bais A or B and B' for 
Rb and Cs.The sign of charge transfer in the ground state of an XY diatomic molecule can be roughly determined by comparing the
asymptotic energy costs $\Delta(XY)=IP(X)-EA(Y)$  versus  $\Delta(YX)=IP(Y)-EA(X)$ describing ionization (IP stands for 
ionization potential) combined with
electron attachement (EA stands for electron affinity) at dissociation. For alkali atoms from Li to Cs, the heavier the alkali 
atom, the smaller the ionization potential (43487, 41449, 35009, 33691, and 31406 cm$^{-1}$)
 and the larger the electron affinity (experimental values are 4984, 4403, 4043, 3020, and  3803 cm$^{-1}$), for 
Li, Na, K, Rb and Cs, respectively \cite{eliav1994}.
It is easy to show that in heteronuclear alkali diatomics including atoms up to Cs, the
electron  transfer from the heavy atom towards the light one is expected to be more favorable, due to
the regular decrease of the IP's with increasing atom size which is not compensated
 by the corresponding variation of  electron affinities. This is indeed confirmed for instance
in our results from paper I, where we chose the convention of a negative sign of the ground state dipole moment in
XY molecules corresponding to a $X^-Y^+$ configuration where the X atom is lighter than the Y atom \cite{aymar2005}.
This model is no longer valid for Fr. Indeed, the ionization potential 
IP(Fr)=32848 cm$^{-1}$ is larger by 1442 cm$^ {-1}$ than the one of Cs, almost half-way 
between IP(Cs) and IP(Rb), whereas the electron affinity EA(Fr)=3920 cm$^{-1}$ 
\cite{eliav2005} (see also ref. \cite{greene1990}) exceeds EA(Cs) by only 117 cm$^{-1}$.
This IP increase  is clearly  related to the decrease of the orbital radius mentionned above.
The energy cost $\Delta(CsFr)$ is lower than $\Delta(FrCs)$ by 1559 cm$^{-1}$, while $\Delta(RbCs)$ is larger
than $\Delta(CsRb)$ by about the same amount. The charge transfer in CsFr is then expected to be Cs$^+$Fr$^-$,
which is confirmed by the positive sign of the dipole moment. The
situation of RbFr is more ambiguous since the difference $\Delta(RbFr)-\Delta(FrRb)=-$57 cm$^{-1}$ is quite small, 
predicting a Rb$^+$Fr$^-$ arrangement.
But the dipole moment is found negative and of small magnitude negative, compatible with a very weak electron 
transfer to form a Rb$^-$Fr$^+$ arrangement.
This apparent disagreement may be due to an uncertainty on the calculated electron-affinity
 of Francium \cite{eliav2005}, but it most probably demonstrates the role of the
polarization forces occurring at finite distance, resulting from
the subtle and self-consistent unscreening of the ion electric field by the one of the electrons.

The equilibrium distance of the lowest triplet state is large, and the state behaves like a neural state
due to spin-forbidden electron transfer. The dipole moment then remains very weak, but we remark that its
variation for RbCs is opposite to the RbFr and RbCs ones.
 At shorter distances, the magnitudes of the dipole moments may be increased due to overlapping
 atomic distributions.

\begin{figure}
\center
\caption{Permanent dipole moment functions of the ground state $(1)^1\Sigma^+$ (panel a) 
and the lowest $(1)^3\Sigma^+$ state (panel b) of RbFr (black dashed line) and CsFr (red dot-dashed line, 
ompared with those of RbCs (blue full line).
A negative dipole moment corresponds to a $X^-Y^+$ configuration.
We used $\alpha_d^{Fr^+}=$20.38~a.u. and basis A for Rb abd Cs.}
\label{fig:permdip}
\end{figure}

\subsection {Excited states of Fr$_2$, RbFr, CsFr and prospects for photoassociation and cold molecule formation}

It is also interesting to look at the Fr$_2$ excited states dissociating to
the $7s+7p$ limits in order to verify if the Fr$_2$ system behaves similarly to the Cs$_2$
and Rb$_2$ species. For instance the existence of long-range molecular wells \cite{stwalley1978} 
has been decisive in the formation process of Cs$_2$ and Rb$_2$ molecules \cite{fioretti1998,gabbanini2000}.
Excited potential curves of $gerade$ and $ungerade$ symmetry correlated to the $7s+7p$ limit
are reproduced in Figure \ref{fig:fr2ex}. Their overall variation is similar to the one for 
all homonuclear alkali pairs, with for instance a crossing at short distances between the
$(1)^1\Sigma_u^+$ and $(1)^3\Pi_u$ curves, which gives rise to an avoided crossing
when spin-orbit interaction is introduced.

The magnitude of the atomic spin-orbit coupling is reported in the
figure, showing that it will be the dominant interaction down to quite short internuclear 
distances ($R \approx 20a_0$). Assuming that a constant spin-orbit interaction is a 
reasonably good approximation for $R>20a_0$ (as it was the case in Cs$_2$, see for 
instance ref.\cite{gutterres2002}), we can diagonalize the electronic Hamiltonian 
including spin-orbit (also given in ref.\cite{gutterres2002}), yielding the curves
reported in Figure \ref{fig:fr2so}. In contrast with the Cs$_2$ \cite{fioretti1999}
and Rb$_2$ \cite{gutterres2002} molecules,
the spin-orbit interaction is so strong in Francium that it dominates the
electrostatic interaction over a range of internuclear distances such that it prevents
the occurence of long-range potential wells in the $0_g^-$ and $1_u$ symmetries, resulting 
respectively from the interaction between the $^3\Pi_g$ and $^3\Sigma_g^+$ states, and between
the $^3\Pi_u$, $^1\Pi_u$, and $^3\Sigma_u^+$ states. For the same reason, the expected avoided crossing
generated by the interaction between the $(1)^1\Sigma_u^+$ and $(1)^3\Pi_u$
states is no more well localized so that the resulting $0_u^+$ curves are almost uncoupled.

\begin{figure}
\center
\caption{Excited potential curves of $gerade$ (panel a) and $ungerade$ (panel c) symmetry 
correlated to the Fr($7s$)+Fr($7p$) dissociation limit, obtained with $\alpha_d^{Fr^+}=$20.38~a.u...
Panel b and d focus on the long-range part of these curves.}
\label{fig:fr2ex}
\end{figure}

\begin{figure}
\center
\caption{Same as Figure \ref{fig:fr2ex}, but including now the fine structure induced by
the atomic spin-orbit interaction.}
\label{fig:fr2so}
\end{figure}

The photoassociation rate for cold Francium atoms can be estimated by comparison with the
cesium one, assuming that similar experimental conditions could be achievable. Following the
model developed in ref.\cite{pillet1997}, we consider a PA transition close to an atomic D
line. The photoassociation rate $R_{PA}$ is then proportional to the product
of characteristic constants $\lambda_{D}^3 \mu^{-1/2} C_3^{2/3} \tau_D^{-1}$, where $\lambda_{D}$
is the wavelength of the D transition, $\mu$ the reduced mass of the atom pair, $C_3$ the effective
long-range coefficient of the PA state, and $\tau_D$ the radiative lifetime of the D line.
Then the ratio of PA rates is estimated according to:

\begin{equation}
\frac{R_{PA}(Fr)}{R_{PA}(Cs)}= \left(\frac{\lambda_{D}(Fr)}{\lambda_{D}(Cs)}\right)^{3}
\left(\frac{\mu_{Cs}}{\mu_{Fr}}\right)^{1/2}\left(\frac{C_3^{Fr}}{C_3^{Cs}}\right)^{2/3}\frac{\tau_D(Cs)}{\tau_D(Fr)}
\label{eq:ratio}
\end{equation}

where the effective $C_3$ coefficient for Francium has been roughly estimated \cite{amiot2002}
from the atomic lifetime of the
D2 line \cite{simsarian1998}. Equation \ref{eq:ratio} results in a factor of about 0.7, meaning that
PA is expected to be almost as efficient in Francium than in Cesium. On the other hand, the absence of
double-well potential, or of a significant interaction between the two $0_u^+$ curves (labelled as 
resonant coupling in ref.\cite{dion2001}) will probably prevent the efficient formation of cold Francium
dimers in deeply bound levels of the ground state or of the lowest triplet state (see refs.
\cite{fioretti1998,dion2001} for a full description of these mechanisms), leaving molecules only in the 
uppermost bound vibrational levels. Moreover, the van der Waals coefficient of ground state Francium 
$C_6^{Fr}=$5226 a.u. is smaller than the one of Cesium $C_6^{Cs}=$6851 a.u.\cite{derevianko2001}.
The corresponding dimer potential curves have a shorter range, which also does not favor the stabilization
of long-range PA molecules by radiative decay towards these states.

The excited potential curves correlated to the two lowest $s+p$ dissociation limits
in RbFr and CsFr are drawn in Figure \ref{fig:mfrex}, as obtained with $\alpha_d^{Fr^+}=$20.38~a.u.,
and with basis B and B' for Rb and Cs. In contrast with all other the mixed pairs, the smallest excitation energy
is now for the lightest atom, which is another characteristic which illustrates the unusual character
of the Francium atom compared to the rest of the alkali series. In RbFr, the spin-orbit is expected to 
strongly mix both $5p+7s$ and $5s+7p$ asymptotes (see the dashed and dotted boxes in Figure \ref{fig:mfrex}, 
which cannot be represented by the perturbative approach
employed for Fr$_2$, and multiple-well structures may probably be expected. The CsFr asymptotes are more spaced 
than in RbFr, so that the perturbative approach may be valid at large distances, but strong mixing will probably 
occur at short interatomic distances.

The van der Waals coefficients for these systems, $C_6^{RbFr}=$4946 a.u. and $C_6^{CsFr}=$5968 a.u.
have the same magnitude than the RbCs one $C_6^{RbCs}=$5663 a.u.\cite{derevianko2001}, so that
the stabilization of PA molecules by radiative decay should be as efficient as in RbCs \cite{kerman2004}.

\begin{figure}
\center
\caption{Excited potential curves for (a) FrRb and (b) FrCs molecules correlated
to the two lowest $s+p$ dissociation limits, obtained with $\alpha_d^{Fr^+}=$20.38~a.u., and with basis B  and B' 
for Rb and Cs respectively. The dashed and dotted area indicate
respectively the atomic fine structure splitting for Rb and Fr
(in panel a), and for Cs and Fr (in panel b). Full lines: singlet states. Dashed lines: triplet states. Black lines:
$\Sigma$ states. Red lines with circles: $\Pi$ states. The energy origin is taken at the $5p+7s$ and $6p+7s$ limit
for RbFr and CsFr respectively.}
\label{fig:mfrex}
\end{figure}

Finally we present in Figure \ref{fig:mfrtran} the transition dipole moments between the ground state and the first excited 
$^1\Sigma^+$ states and $^1\Pi$ in Francium compounds. Here also results are equivalent 
when using basis A or basis B and B' for Rb and Cs respectively. In the upper panel, one
sees that the RbFr and CsFr dipole moments for
the transition towards the lowest $\Sigma$ excited state are superimposed onto the 
Fr$_2$ function between 7 and 12$a_0$, suggesting 
that in this range, the electrons are closer to the Francium cores confirming the discussion about permanent dipole moments above. The same
pattern is observable in panel b as well, but over a shorter range (8 to 10 $a_0$). The presence of a strong avoided crossing between the
$^1\Pi$ states in RbFr and CsFr is visible here as the two transition dipole moment
functions cross each other at the position of the avoided crossing.

\begin{figure}
\center
\caption{Transition dipole moments from the ground state towards the first excited (a) $^1\Sigma^+$ states and (b)
$^1\Pi$ in Francium compounds. Full black lines: $X \rightarrow (1)^1\Sigma_u^+ (7s+7p)$ and $X \rightarrow (1)^1\Pi_u (7s+7p)$
in Fr$_2$. Full red lines with closed circles: $X \rightarrow (1)^1\Sigma^+ (5p+7s)$ and $X \rightarrow (1)^1\Pi^(5p+7s)$ in
RbFr. Dashed red lines with closed circles: $X \rightarrow (2)^1\Sigma^+ (5s+7p)$ and $X \rightarrow (2)^1\Pi (5s+7p)$ in
RbFr. Full blue lines with open circles: $X \rightarrow (1)^1\Sigma^+ (6p+7s)$ and $X \rightarrow (1)^1\Pi (6p+7s)$ in
CsFr. Dashed blue lines with open circles: $X \rightarrow (2)^1\Sigma^+ (6s+7p)$ and $X \rightarrow (2)^1\Pi (6s+7p)$ in
CsFr.Calculations are performed with $\alpha_d^{Fr^+}=$20.38~a.u., and with basis A for
both Rb and Cs.}
\label{fig:mfrtran}
\end{figure}

\section{Conclusion}

In this work we investigated the possibility to create cold Fr$_2$, RbFr, and CsFr molecules
through photoassociation of cold atoms. Indeed, the creation of dense samples of cold
Francium atoms may become available experimentally, as well as the mixing with cold 
Rubidium or Cesium atoms. We predict that under comparable experimental conditions, 
photoassociation of cold Francium atoms should be as efficient as with cold cesium atoms.
In contrast the formation of ultracold Fr$_2$  molecules will probably be weaker as the formation
processes known in Cs$_2$, relying on the presence of long-range potential wells or 
resonant coupling, are not expected in Fr$_2$. However, the formation of ultracold RbFr 
or CsFr molecules is probably as efficient as in RbCs, which could be worthwhile to try
experimentally as dense cold Rb and Cs samples are widely available.

To extract these results, we performed the first calculations of the electronic structure
of the Fr$_2$, RbFr, and CsFr molecules, yielding potential curves, permanent and
transition dipole moments. A new scalar (ie spin-orbit averaged) pseudopotential has
been determined to represent the interaction of the valence electron with the large 
Fr$^+$ core complemented with core-polarization operators. Through a comparison with the
closest heavy alkali dimers Rb$_2$, Cs$_2$ and RbCs, we found that our results yield a 
realistic representation of the Fr$_2$, FrCs and FrRb molecules.  It would be interesting
to check the present results with  two-component pseudopotential configuration 
interaction schemes \cite{ermler2002}, but they only exist in the framework of
small core pseudopotential (ie 10 valence electrons or more in he case of Francium. This 
would allow in particular to estimate errors of the present calculation which ignore
spin-orbit induced polarization effects. However, we provided a description of the molecular 
fine structure via an empirical atom-in molecules schemes, which suggests
that the behaviour of excited Fr$_2$ molecular states involved in photoassociation and
cold molecule formation is different from Rb$_2$ and Cs$_2$ cases. Here again, it would be 
interesting to compare with more refined ab initio spin-orbit calculations which are
in progress. An interesting conclusion of the present work is that, while the nature and 
the quantitative behaviour of the ground state and metastable triplet state potential
curves of the francium compounds are generally comparable with other previously 
studied alkali diatomics, the ground state polarity of hetero-atomic molecules 
cannot be extrapolated from the
other alkali pairs, due to the influence of relativistic effect which affect atomic  properties, 
especially the ionization potential(correlated with the outer $s$ orbital radius).
We hope that the present work will stimulate experimental spectroscopic studies on those
short-lived species, which would also help to  discuss further the accuracy of the present results.

\ack
This work has been performed in the framework of the Research and Training Network of
the European Union "Cold Molecules" (contract number HPRN-CT-2002-00290). Discussions
with P. Indelicato about the Dirac-Fock program are gratefully acknowledged.

\section*{References}

\end{document}